\documentstyle[aps,eqsecnum]{revtex}

\begin{document}
\title{On Hamiltonian formulations of the Schr\"{o}dinger system}
\author{L\'{a}szl\'{o} \'{A}. Gergely}
\address{Astronomical Observatory and Department of Experimental Physics,\\
University of Szeged, D\'{o}m t\'{e}r 9, Szeged, H-6720 Hungary}
\maketitle

\begin{abstract}
We review and compare different variational formulations for the Schr\"{o}%
dinger field. Some of them rely on the addition of a conveniently chosen
total time derivative to the hermitic Lagrangian. Alternatively, the
Dirac-Bergmann algorithm yields the Schr\"{o}dinger equation first as a
consistency condition in the full phase space, second as canonical equation
in the reduced phase space. The two methods lead to the same (reduced)
Hamiltonian. As a third possibility, the Faddeev-Jackiw method is shown to
be a shortcut of the Dirac method. By implementing the quantization scheme
for systems with second class constraints, inconsistencies of previous
treatments are eliminated.
\end{abstract}

\section{Introduction}

Outstanding equation of modern physics, the Schr\"{o}dinger equation has
multiple and deep connections with integral principles. Historically, Schr%
\"{o}dinger obtained his equation guided by the beautiful analogy between
the Fermat principle and the principle of least action \cite{Schrod}.
Moreover, motivated by a remark of Dirac \cite{Dirac1}, Feynman has derived
the Schr\"{o}dinger equation from the Huygens principle, realizing the first
step towards his path integral approach \cite{Feynman1}, \cite{Feynman2}.
(For recent recent reviews see \cite{Tzanakis}, \cite{Derbes}.) The Schr\"{o}%
dinger field is equally a popular choice to illustrate how second
quantization proceeds.

At a closer inspection, however, the power and beauty of the variational
approach is obstructed by an aesthetical bug. The reason: the abundance of
dynamical variables, some of them being redundant. The Lagrangian yielding
the nonrelativistic Schr\"{o}dinger equation is linear in the time
derivatives of the fields $\Psi $ and its complex conjugate $\Psi ^{\ast }$.
Thus the Legendre transformation does not lead to an unambiguous
Hamiltonian. Various Hamiltonians, all having Schr\"{o}dinger's equation as
canonical equation can be found in textbooks. They depend either on two
pairs of canonical variables \cite{HT}, or - as a result of adding a total
time derivative to the hermitic Lagrangian - just on a single (complex or
real) canonical pair \cite{Schiff}-\cite{CT}. In the next section we briefly
rewiev these approaches, pointing out both the weakness and ingenuity in
sweeping away the problem.

There are two equivalent methods to circumvent this difficulty. In section 3
we treat the Schr\"{o}dinger field as a constrained system, applying the
Dirac-Bergmann algorithm \cite{Dirac2}-\cite{Sund}. As the system has second
class constraints, the dynamics involves Dirac brackets. We present two
alternatives to the existing derivations of the Schr\"{o}dinger equation.
First, the consistency requirement yields the Schr\"{o}dinger equation in
the form of a weak equation on the full phase space. Second, by a suitable
canonical transformation we introduce new canonical coordinates, containing
the constraints. The Dirac bracket of the full phase space becomes the
Poisson bracket of the reduced phase space and the Schr\"{o}dinger equation
is a canonical equation. As a bonus we recover the Hamiltonian obtained by
addition of a properly chosen time-derivative. At the end of Section 3 we
present an alternative discussion in terms of real variables: the real and
imaginary parts of the complex field $\Psi $.

In Section 4 we apply the Faddeev-Jackiw scheme developed for Lagrangians,
which are first order in ''velocities'' \cite{Jackiw1}, \cite{Jackiw2}. We
verify that the fundamental brackets of the Faddeev-Jackiw approach coincide
with the Dirac brackets.

Finally in the fifth section we follow the canonical quantization scheme %
\cite{Dirac2}-\cite{Sund}, \cite{Dirac3}, giving an operator representation
of the Dirac bracket algebra of the canonical variables. This is equivalent
with the quantization of the Faddeev-Jackiw fundamental bracket. Our
approach avoids the interpretational inconsistencies of some standard
treatments \cite{HT}, \cite{Schiff}, already pointed out by Tassie \cite%
{Tassie} and is a viable alternative to the existing reduced phase space
quantization schemes \cite{Kuchar}, \cite{CT}. By imposing the second class
constraints as operator identities, second quantization proceeds smoothly.

\section{Standard variational procedures. A review}

\subsection{Two pairs of complex variables}

The action for the Schr\"{o}dinger field cf. Henley and Thirring \cite{HT}
is: 
\begin{eqnarray}
{S[\psi ,\psi ^{\ast }]\ } &=&{\ \int dt\int d{\bf r}\ {\cal L}\ }
\label{actionHT} \\
{{\cal L}\ } &=&{\ {\frac{i\hbar }{2}}\left( \psi ^{\ast }\dot{\psi}-\psi 
\dot{\psi}^{\ast }\right) -{\frac{\hbar ^{2}}{2m}}\nabla \psi ^{\ast }\nabla
\psi -V({\bf r},t)\psi ^{\ast }\psi \ .}  \nonumber
\end{eqnarray}%
This Lagrangian density ${\cal L}$ is hermitian\footnote{%
The Lagrangian density (\ref{actionHT}) when written in terms of the modulus
and argument of $\psi $, leads to equations very similar to the equations of
hydrodynamical flow in presence of a potential $V$. This analogy stands at
the base of the hydrodynamical model of quantum mechanics.}. Variation of (%
\ref{actionHT}) with respect to $\psi ^{\ast }$ and $\psi $ gives the Schr%
\"{o}dinger equation 
\begin{equation}
{i\hbar \dot{\psi}+{\frac{\hbar ^{2}}{2m}}\Delta \psi -V\psi \ }={}{\ 0}\ ,
\label{Schrodeq}
\end{equation}%
and its complex conjugate. The canonical momenta of the complex conjugate
variables are complex conjugates too: 
\begin{equation}
\pi \ =\ {\frac{i\hbar }{2}}\psi ^{\ast }\ ,\qquad \pi ^{\ast }\ =\ -{\frac{%
i\hbar }{2}}\psi \ .  \label{momentaHT}
\end{equation}%
According to Henley and Thirring, the Hamiltonian density ${\cal H}$, when
written in terms of the phase space variables $(\psi ,\ \psi ^{\ast },\ \pi
,\ \pi ^{\ast })$, {\it has} to be: 
\begin{eqnarray}
{S[\psi ,\pi ,\psi ^{\ast },\pi ^{\ast }]\ } &=&{\ \int dt\int d{\bf r}\
(\pi \dot{\psi}+\pi ^{\ast }\dot{\psi}^{\ast }-{\cal H})\ }  \nonumber \\
{{\cal H}\ } &=&{\ {\frac{i\hbar }{2m}}\left( \nabla \psi ^{\ast }\nabla \pi
^{\ast }-\nabla \psi \nabla \pi \right) -{\frac{i}{\hbar }}V({\bf r},t)(\psi
\pi -\psi ^{\ast }\pi ^{\ast })\ .}  \label{actionHTham}
\end{eqnarray}%
Variations with respect to $\psi ,\ \psi ^{\ast }$ give the Schr\"{o}dinger
equation and its complex conjugate. The same equations emerge as a result of
the relations (\ref{momentaHT}) and the variations with respect to $\pi ,\
\pi ^{\ast }$. The Schr\"{o}dinger equation as such a frequent outcome
indicates that there are redundant variables in the formalism.

Note that it is impossible to express the velocities $\dot{\psi},\ \dot{\psi}%
^{\ast }$ in terms of canonical data from the expressions of momenta (\ref%
{momentaHT}), as would be required by the Legendre transformation. How can
one then find the Hamiltonian (\ref{actionHTham})? No hint for this is given
in \cite{HT}. We can still invent a method. First express the fields $\psi
,\ \psi ^{\ast }$ from (\ref{momentaHT}), insert them in the {\it kinetic}
terms of (\ref{actionHT}), but not in the time derivatives. Now the action (%
\ref{actionHT}) takes an {\it already Hamiltonian form}, similarly as (\ref{actionHTham}).
At this point we see a Hamiltonian expressed in terms of canonical data,
which is hermitian, however does not give the correct equations. Instead we
devide the potential terms of (\ref{actionHT}) in two equal parts, eliminate
the starred fields from one part and the unstarred ones from the other part
by means of the relations (\ref{momentaHT}). The hermiticity of the
Hamiltonian is kept and we get the Schr\"{o}dinger equation. Proceeding in
other ways, for example, eliminating all starred or all unstarred fields
(and consequently destroying the hermiticity) will result in wrong
Hamiltonians. However, as we have seen, hermiticity alone is not a criteria
for correctness.

\subsection{One pair of complex variables}

Furthermore, the requirement of hermiticity is not even a necessary one.
Schiff \cite{Schiff} derives the Schr\"{o}dinger equation from a
non-hermitic Lagrangian, found from (\ref{actionHT}) by adding the total
time derivative $i\hbar {(\psi \psi ^{\ast })}^{.}/2$: 
\begin{eqnarray}
{S[\psi ,\psi ^{\ast }]\ } &=&{\ \int dt\int d{\bf r}\ {\cal L_{S}}\ } 
\nonumber \\
{{\cal L_{S}}\ } &=&{\ i\hbar \psi ^{\ast }\dot{\psi}-{\frac{\hbar ^{2}}{2m}}%
\nabla \psi ^{\ast }\nabla \psi -V({\bf r},t)\psi ^{\ast }\psi \ .}
\label{actionSchiff}
\end{eqnarray}%
Variation of the action (\ref{actionSchiff}) with respect to $\psi $ and $%
\psi ^{\ast }$ gives the Schr\"{o}dinger equation (\ref{Schrodeq}) and its
complex conjugate. However the canonical momenta are not complex conjugate
any more: 
\begin{equation}
\pi _{S}\ :=\ {\frac{\delta L_{S}}{\delta \dot{\psi}}}\ =\ i\hbar \psi
^{\ast }\ ,\qquad \pi _{S}^{\ast }\ :=\ {\frac{\delta L_{S}}{\delta \dot{\psi%
}^{\ast }}}\ =0\ .  \label{momentaSchiff}
\end{equation}%
The first of these equations is used to eliminate $\psi ^{\ast }$ from the
action, which becomes: 
\begin{eqnarray}
{S[\psi ,\pi _{S}]\ } &=&{\ \int dt\int d{\bf r}\ (\pi _{S}\dot{\psi}-{\cal %
H_{S}})\ }  \nonumber \\
{{\cal H_{S}}\ } &=&{\ -{\frac{i\hbar }{2m}}\nabla \psi \nabla \pi _{S}-{\ 
\frac{i}{\hbar }}V({\bf r},t)\psi \pi _{S}\ .}  \label{actionSchiffham}
\end{eqnarray}%
Thus no starred field shows up as canonical variable. The canonical equation
obtained by varying $\psi $ is the Schr\"{o}dinger equation. The other
canonical equation from variation of $\pi _{S}$, together with the first
relation (\ref{momentaSchiff}) gives the complex conjugate Scr\"{o}dinger
equation. Some of the superfluous variables were removed by the addition of
a total time derivative, achieving a partial reduction to the true degrees
of freedom. This Hamiltonian description is the simplest one in terms of
complex fields.

\subsection{One pair of real variables}

A more efficient way to look on the variational problem for quantum
mechanics is described by Kucha\v{r} \cite{Kuchar} and alternatively by
Cohen-Tannoudji, Dupont-Roc and Grynberg \cite{CT}. These approaches rely on
the decomposition of the field $\psi $ in real and imaginary parts: 
\begin{equation}
q={\frac{1}{\sqrt{2}}}(\psi +\psi ^{\ast })\ ,\qquad p=-{\frac{i\hbar }{%
\sqrt{2}}}(\psi -\psi ^{\ast })\ .  \label{qp}
\end{equation}%
In terms of the real fields $q$ and $p$ the action (\ref{actionHT}) becomes: 
\begin{eqnarray}
{S[q,p]\ } &=&{\ \int dt\int d{\bf r}\ {\cal L}\ }  \nonumber \\
{{\cal L}\ } &=&{\ {\frac{1}{2}}(p\dot{q}-q\dot{p})-{\frac{\hbar ^{2}}{4m}}%
\left[ (\nabla q)^{2}+\left( {\frac{\nabla p}{\hbar }}\right) ^{2}\right] -{%
\ \frac{V}{2}}\left( q^{2}+{\frac{p^{2}}{\hbar ^{2}}}\right) \ .}
\label{actionK}
\end{eqnarray}%
By adding the total time derivative ${(pq)}^{.}/2$ to the Lagrangian
density, the action takes an already Hamiltonian form. The field $p$ turns
to be the conjugate momentum to $q$: 
\begin{eqnarray}
{S[q,p]\ } &=&{\ \int dt\int d{\bf r}\ (p\dot{q}-{\cal H_{K}})\ }  \nonumber
\\
{{\cal H_{K}}\ } &=&{\ {\frac{\hbar ^{2}}{4m}}\left[ (\nabla q)^{2}+\left( {%
\ \frac{\nabla p}{\hbar }}\right) ^{2}\right] +{\frac{V}{2}}\left( q^{2}+{\ 
\frac{p^{2}}{\hbar ^{2}}}\right) \ .}  \label{actionKham}
\end{eqnarray}%
The two canonical equations obtained by variations with respect to $q$ and $p
$ are: 
\begin{equation}
\dot{q}=-{\frac{1}{2m}}\Delta p+{\frac{V}{\hbar ^{2}}}p\ ,\qquad \dot{p}={\ 
\frac{\hbar ^{2}}{2m}}\Delta q-Vq\ .  \label{Schrodreim}
\end{equation}%
Up to global factors, the first equation of (\ref{Schrodreim}) is the real
part, while the second the imaginary part of the Schr\"{o}dinger equation (%
\ref{Schrodeq}).

In this latest approach a full reduction of the phase space to the true
degrees of freedom was achieved, as the Schr\"odinger equation emerges only
once in the Hamiltonian formalism. In terms of real fields this is the
simplest description, which again relies on the addition of a properly
chosen total time derivative to the Lagrangian density.

We will see in the next section that the addition of specific total time
derivative terms to the Lagrangian (\ref{actionHT}) is not compulsory. The
standard Dirac-Bergmann algorithm leads directly either to the Hamiltonian
density (\ref{actionSchiffham}), (in a description in terms of complex
fields) or the Hamiltonian density (\ref{actionKham}) (if real fields are
introduced).

\section{The Dirac-Bergmann algorithm}

Constrained systems are characterized by the singularity of the inertia
matrix, whoose elements are given by the second derivatives of the
Lagrangian with respect to the generalized velocities. This property always
holds if a Lagrangian density is linear in the time derivatives of fields %
\cite{Sund}, as is the case for (\ref{actionHT}). Then the Dirac-Bergmann
algorithm takes the role of the Legendre transformation. 

\subsection{Complex fields}

The momenta (\ref{momentaHT}) provide two primary Hamiltonian constraints: 
\begin{equation}
\phi _{1}\ :=\ \pi -{\frac{i\hbar }{2}}\psi ^{\ast }\ =\ 0\ ,\qquad \phi
_{2}\ :=\ \pi ^{\ast }+{\frac{i\hbar }{2}}\psi \ =\ 0\ .  \label{primary}
\end{equation}%
Time evolution of an arbitrary phase space function  $f$ is generated
through the Poisson bracket by the primary Hamiltonian density ${\cal H_{P}}$
rather than the canonical Hamiltonian density ${\cal H_{C}}$ : 
\begin{eqnarray}
{\dot{f}\ } &\approx &{\{f,H_{P}\}\ ,\qquad H_{P}=\int d{\bf r}\ {\cal H_{P}}%
}  \nonumber \\
{{\cal H_{P}}\ :=} &&{\ {\cal H_{C}}+\dot{\psi}\phi _{1}+\dot{\psi}^{\ast
}\phi _{2}}  \nonumber \\
{{\cal H_{C}}\ :=} &&{\ {\frac{\hbar ^{2}}{2m}}\nabla \psi ^{\ast }\nabla
\psi +V({\bf r},t)\psi ^{\ast }\psi \ .}  \label{timevol}
\end{eqnarray}%
Here the velocities $\dot{\psi},\ \dot{\psi}^{\ast }$ are unknown functions
of the phase space variables. The symbol $\approx $ denotes weak equality,
holding only on the hypersurface determined by the constraints (\ref{primary}%
).

Consistency requires the time derivatives of the primary constraints $\phi
_{1},\ \phi _{2}$ to vanish: 
\begin{eqnarray}
{0\ } &\approx &\dot{\phi}_{1}\approx -i\hbar \dot{\psi}^{\ast }+{\frac{%
\hbar ^{2}}{2m}}\Delta \psi ^{\ast }-V\psi ^{\ast }  \nonumber \\
{0\ } &\approx &\dot{\phi}_{2}\approx i\hbar \dot{\psi}+{\frac{\hbar ^{2}}{2m%
}}\Delta \psi -V\psi ^{\ast }\ .  \label{Schrodeqweak}
\end{eqnarray}%
These are the Schr\"{o}dinger equation and its complex conjugate. They
emerge as weak equalities; a price one has to pay for working on the
complete phase space ($\psi ,\ \pi ,\ \psi ^{\ast },$ $\pi ^{\ast })$. No
secondary constraint appears in the theory since Eqs. (\ref{Schrodeqweak})
are relations determining the unknown functions $\dot{\psi}$ and $\dot{\psi}%
^{\ast }$.

The Poisson bracket of the two constraints shows that they are of second
class: 
\begin{equation}
\{\phi _{1},\phi _{2}\}\ :=\{\phi _{1}({\bf r},t),\phi _{2}({\bf r^{\prime }}%
,t)\}\ =\ -i\hbar \delta ({\bf r}-{\bf r^{\prime }})\ .  \label{secondclass}
\end{equation}

Time evolution can be given in the alternative form: 
\begin{equation}
\dot{f}\ =\ \{f,H_{C}\}_{D}\ ,\qquad H_{C}=\int d{\bf r}\ {\cal H_{C}}
\label{timevolD}
\end{equation}%
in terms of the Dirac bracket \cite{Dirac2}: 
\begin{equation}
\{f,g\}_{D}\ :=\ \{f,g\}-\sum_{i,j=1,2}\{f,\phi _{i}\}\left( \{\phi
_{k},\phi _{l}\}\right) _{ij}^{-1}\{\phi _{j},g\}\ .  \label{DB}
\end{equation}%
Here $\left( \{\phi _{k},\phi _{l}\}\right) ^{-1}$ denotes the inverse of
the matrix with elements given by the Poisson brackets of the constraints.
Straightforward computation using (\ref{secondclass}) gives the following
expression for the Dirac bracket: 
\begin{equation}
\{f,g\}_{D}\ :=\ {\frac{1}{2}}\{f,g\}-{\frac{i}{\hbar }}\int d{\bf r}\left( {%
\ \frac{\delta f}{\delta \psi }}{\frac{\delta g}{\delta \psi ^{\ast }}}-{%
\frac{\delta f}{\delta \psi ^{\ast }}}{\frac{\delta g}{\delta \psi }}\right)
+{\frac{i\hbar }{4}}\int d{\bf r}\left( {\frac{\delta f}{\delta \pi }}{\frac{%
\delta g}{\delta \pi ^{\ast }}}-{\frac{\delta f}{\delta \pi ^{\ast }}}{\frac{%
\delta g}{\delta \pi }}\right)   \label{DBexpl}
\end{equation}%
From here it is immediate to write the Dirac brackets of the canonical data: 
\begin{equation}
\{\psi ({\bf r}),\psi ^{\ast }({\bf r^{\prime }})\}_{D}\ =\ -{\frac{i}{\hbar 
}}\delta ({\bf r}-{\bf r^{\prime }})\ ,\qquad \{\psi ({\bf r}),\psi ({\bf %
r^{\prime }})\}_{D}\ =\ \{\psi ^{\ast }({\bf r}),\psi ^{\ast }({\bf %
r^{\prime }})\}_{D}\ =\ 0\ ,  \label{DB1}
\end{equation}%
\begin{equation}
\{\pi ({\bf r}),\pi ^{\ast }({\bf r^{\prime }})\}_{D}\ =\ {\frac{i\hbar }{4}%
}\delta ({\bf r}-{\bf r^{\prime }})\ ,\qquad \{\pi ({\bf r}),\pi ({\bf %
r^{\prime }})\}_{D}\ =\ \{\pi ^{\ast }({\bf r}),\pi ^{\ast }({\bf r^{\prime }%
})\}_{D}\ =\ 0\ ,  \label{DB2}
\end{equation}%
\begin{equation}
\{\psi ({\bf r}),\pi ^{\ast}({\bf %
r^{\prime }})\}_{D}\ =\ \{\psi ^{\ast }({\bf r}),\pi ({\bf r^{\prime }%
})\}_{D}\ =\ 0\ ,  \label{DB2a}
\end{equation}%
\begin{equation}
\{\psi ({\bf r}),\pi ({\bf r^{\prime }})\}_{D}\ =\ {\frac{1}{2}}\{\psi ({\bf %
r}),\pi ({\bf r^{\prime }})\}\ =\ {\frac{1}{2}}\delta ({\bf r}-{\bf %
r^{\prime }})\ ,  \label{DB3}
\end{equation}%
\begin{equation}
\{\psi ^{\ast }({\bf r}),\pi ^{\ast }({\bf r^{\prime }})\}_{D}\ =\ {\frac{1}{%
2}}\{\psi ^{\ast }({\bf r}),\pi ^{\ast }({\bf r^{\prime }})\}\ =\ {\frac{1}{2%
}}\delta ({\bf r}-{\bf r^{\prime }})\ .  \label{DB4}
\end{equation}%
These Dirac brackets do not contain phase-space functions, thus no operator
ordering difficulties will occur during quantization.

{\bf Reduction.} Dirac brackets of second class constraints with arbitrary
functions vanish. Thus the constraints can be solved prior to calculating
the Dirac brackets, by reducing the phase space to the physical degrees of
freedom. As each second class constraint reduces the dimension of the phase
space by one, a basis in the reduced phase space is provided by a single
pair of complex canonical data. A suitable canonical transformation turns
the constraints into the other pair of canonical data: 
\begin{equation}
\left( 
\begin{array}{ll}
{\psi } & \qquad {\psi ^{\ast }} \\ 
{\pi } & \qquad {\pi ^{\ast }}%
\end{array}%
\right) \rightarrow \left( 
\begin{array}{ll}
{\psi _{1}=\psi /2+}i{\pi ^{\ast }/\hbar } & \qquad {\psi _{2}=-}i{\phi
_{2}/\hbar } \\ 
{\pi _{1}=\pi +i\hbar \psi ^{\ast }/2} & \qquad {\pi _{2}=\phi _{1}}%
\end{array}%
\right)   \label{chart1}
\end{equation}%
A straightforward check shows that the Dirac bracket (\ref{DBexpl}) written
in terms of the new coordinates becomes the Poisson bracket of the reduced
phase space, coordinatized by $\psi _{1},\ \pi _{1}$: 
\begin{equation}
\{f,g\}_{D}\ =\ \int d{\bf r}\left( {\frac{\delta f}{\delta \psi _{1}}}{\ 
\frac{\delta g}{\delta \pi _{1}}}-{\frac{\delta f}{\delta \pi _{1}}}{\frac{%
\delta g}{\delta \psi _{1}}}\right) \ .  \label{DBPB}
\end{equation}%
Now it is immediate to verify the generic property that the Dirac brackets
of the constraints $\psi _{2}$ and $\pi _{2}$ with arbitrary functions
vanish.

The Hamiltonian density on the reduced phase space is the one introduced by
Schiff (\ref{actionSchiffham}), with $\psi _{1},\ \pi _{1}$ in place of $%
\psi $ and $\pi _{S}$. The canonical equations are the Schr\"{o}dinger
equation and its complex conjugate.

\subsection{Real fields}

Our starting point in this section is the action (\ref{actionK}) written in
terms of the real fields $q$ and $p$. As in the previous section we have
presented in detail the method, here we merely list the results. From the
definition of the momenta $P_{q,p}$ canonically conjugate to the $q,p$
variables we find the constraints which are second class: 
\begin{eqnarray}
\Phi _{1}:=P_{q}\  &-&\ {\frac{p}{2}}\ ,\qquad \Phi _{2}:=P_{p}\ +\ {\frac{q%
}{2}}\ ,  \nonumber \\
\{\Phi _{2},\Phi _{1}\}:= &&\delta ({\bf r}-{\bf r^{\prime }})\ .
\label{seclassconstr}
\end{eqnarray}%
The consistency requirements $\dot{\Phi}_{1,2}\ \approx \ 0$ are the real
and imaginary parts of the Schr\"{o}dinger equation, Eqs. (\ref{Schrodreim}).

{\bf Reduction.} The constraints (\ref{seclassconstr}) already form a
canonical pair of variables, thus a canonical transformation to the reduced
phase space is immediate: 
\begin{equation}
\left( 
\begin{array}{ll}
q & \qquad p \\ 
{P_{q}} & \qquad {P_{p}}%
\end{array}%
\right) \rightarrow \left( 
\begin{array}{ll}
Q_{1}=q/2-P_{p} & \qquad Q_{2}=\Phi _{2} \\ 
P_{1}=p/2+P_{q} & \qquad P_{2}={\Phi _{1}}%
\end{array}%
\right)   \label{chart2}
\end{equation}%
Again the Dirac bracket on the full phase space ($Q_{1,2},P_{1,2})$ becomes
the Poisson bracket on the reduced phase space ($Q_{1},P_{1})$. The
Hamiltonian density on this reduced phase space is ${\cal H_{K}}$ given in (%
\ref{actionKham}), with $Q_{1},P_{1}$ in place of $q,p$. So the canonical
equations on the reduced phase space are the real and imaginary parts of the
Schr\"{o}dinger equation.

\section{The Faddeev-Jackiw approach}

Developed as a Hamiltonian formulation of dynamical systems with Lagrangians
linear in velocities, the Faddeev-Jackiw method provides undoubtfully the
shortest path toward a fundamental bracket in the phase space.

The kinetic part of the Lagrangian (\ref{actionHT}) with $2\times \infty $
basic variables $\xi ^{i}({\bf r})=(\psi ({\bf r^{\prime }}),\ \psi ^{\ast }(%
{\bf r"}))$ determines the symplectic $2$-form with the
inverse 
\begin{equation}
\omega ^{ij}({\bf r^{\prime }},{\bf r}^{\prime \prime })=\frac{1}{i\hbar }%
\left( \matrix{0 & \delta({\bf r'}-{\bf r''})\cr -\delta({\bf r'}-{\bf r''})
& 0}\right) \ ,
\end{equation}%
in terms of which the time evolution of the fundamental variables is 
\begin{equation}
\dot{\xi}^{i}({\bf r})=\omega ^{ij}({\bf r},{\bf r^{\prime }})\frac{\delta }{%
\delta \xi ^{j}({\bf r^{\prime })}}H_{C}\ .  \label{FJ}
\end{equation}%
This is nothing but the Schr\"{o}dinger equation and its complex conjugate.
Eq. (\ref{FJ}) represents a Hamiltonian evolution if the fundamental bracket
obeys 
\begin{equation}
\{\xi ^{i}({\bf r}),\xi ^{j}({\bf r^{\prime }})\}_{FJ}=\omega ^{ij}({\bf r},%
{\bf r^{\prime }})\ ,
\end{equation}%
which is nothing but a shorthand notation for Eqs. (\ref{DB1}).

\section{Second quantization}

The standard procedure for canonical quantization of systems with second
class constraints is to turn the Dirac brackets into commutators cf. the
scheme: 
\begin{equation}
\{f,g\}_{D}=l\qquad \rightarrow \qquad \lbrack \hat{f},\hat{g}]=i\hbar \hat{l%
}.  \label{qscheme}
\end{equation}%
Here $f,g$ and $l$ are phase-space functions, $\hat{f},\hat{g}$ and $\hat{l}$
are operators. The complex conjugate $f^{\ast }$ of a function $f$ becomes
the adjoint operator $\hat{f}^{\dagger }$. When we apply these prescriptions
to the Dirac brackets (\ref{DB1})-(\ref{DB4}), we get: 
\begin{equation}
\lbrack \hat{\psi}({\bf r}),\hat{\psi}^{\dagger }({\bf r^{\prime }})]\ =\
\delta ({\bf r}-{\bf r^{\prime }})\ ,\qquad \lbrack \hat{\psi}({\bf r}),\hat{%
\psi}({\bf r^{\prime }})]\ =\ [\hat{\psi}^{\dagger }({\bf r}),\hat{\psi}%
^{\dagger }({\bf r^{\prime }})]\ =\ 0\ ,  \label{comm1}
\end{equation}%
\begin{equation}
\lbrack \hat{\pi}({\bf r}),\hat{\pi}^{\dagger }({\bf r^{\prime }})]\ =\ -{\ 
\frac{\hbar ^{2}}{4}}\delta ({\bf r}-{\bf r^{\prime }})\ ,\qquad \lbrack 
\hat{\pi}({\bf r}),\hat{\pi}({\bf r^{\prime }})]\ =\ \{\hat{\pi}^{\dagger }(%
{\bf r}),\hat{\pi}^{\dagger }({\bf r^{\prime }})]\ =\ 0\ ,  \label{comm2}
\end{equation}%
\begin{equation}
\lbrack \hat{\psi}({\bf r}),\hat{\pi}({\bf r^{\prime }})]\ =\ [\hat{\psi}%
^{\dagger }({\bf r}),\hat{\pi}^{\dagger }({\bf r^{\prime }})]\ =\ {\frac{%
i\hbar }{2}}\delta ({\bf r}-{\bf r^{\prime }})
\ ,\qquad
\lbrack \hat{\psi}({\bf r}),\hat{\pi}^{\dagger }({\bf r^{\prime }})]\ 
=\ [\hat{\psi}%
^{\dagger }({\bf r}),\hat{\pi}({\bf r^{\prime }})]\ =\ 0
\ .  \label{comm3}
\end{equation}%
The second class constraints (\ref{primary}) of the theory become operator
identities: 
\begin{equation}
\hat{\pi}={\frac{i\hbar }{2}}\hat{\psi}^{\dagger }\ ,\qquad \hat{\pi}%
^{\dagger }=-{\frac{i\hbar }{2}}\hat{\psi}\ .  \label{opmomenta}
\end{equation}%
By inserting Eqs. (\ref{opmomenta}) in Eqs. (\ref{comm2}) and (\ref{comm3})
we recover again Eqs. (\ref{comm1}).

The commutators (\ref{comm1}) between $\hat{\psi}$ and $\hat{\psi}^{\dagger }
$ represent the starting point in the second quantization, as indicated by
Henley and Thirring \cite{HT} and described in detail by Schiff \cite{Schiff}%
. While Eqs. (\ref{comm1}) have emerged in a natural way from the Dirac
bracket quantization, they had to be imposed ''by hand'' in the previous
approaches, a feature already criticized by Tassie \cite{Tassie}. Schiff arrives to
Eqs. (\ref{comm1}) by imposing the canonical commutation rules: 
\begin{equation}
\lbrack \hat{\psi},\hat{\pi}_{S}]\ =\ [\hat{\psi}^{\dagger },\hat{\pi}%
_{S}^{\dagger }]\ =\ i\hbar \delta ({\bf r}-{\bf r^{\prime }}),
\label{cancomm}
\end{equation}%
however his treatment also requires $\hat{\pi}_{S}^{\dagger }=0$ as can be
seen from Eq. (\ref{momentaSchiff}), which is in obvious contradiction with
the second commutator (\ref{cancomm}). Meanwhile, the treatment of Henley
and Thirring starts by postulating Eqs. (\ref{comm3}), in other words with
the surprizing statement that what was canonically conjugate in the
classical theory in not any more canonically conjugate in the quantum theory
(see the extra factor of $1/2$). They impose the commutators (\ref{comm3})
motivated by the analogy with the variational problem of the harmonic
oscillator, written in complex coordinates. Tassie proposes a solution to
these conceptual problems. By working on the momentum space, he esentially
eliminates the imaginary part of $\psi $ and gives a description in terms of
real fields without encountering the above-mentioned inconvenience.

No problems appear in the approaches employing real fields. This is because
any description in terms of real fields \cite{Kuchar,CT,Tassie} esentially
means that we have eliminated the redundant variables, thus we are
quantizing on the reduced phase-space.

No difficulties appear in the Dirac bracket quantization either. The Dirac
bracket (\ref{DB3}) of the variables $\psi $ and $\pi $, canonically
conjugate at the classical level, is one-half their Poisson bracket, thus no
reason to wonder why the corresponding commutator contains the factor of $1/2
$. Consistent canonical quantization of the complex Schr\"{o}dinger field
requires the Dirac bracket.

Stated in other way, if we start from the canonical chart (\ref{chart1}),
compute the Dirac bracket cf. Eq. (\ref{DBPB}) and apply the prescription (%
\ref{qscheme}) for the variables spanning the reduced phase space, we find: 
\begin{equation}
\{\psi _{1}({\bf r}),\pi _{1}({\bf r^{\prime }})\}_{D}\ =\ \delta ({\bf r}-%
{\bf r^{\prime }})\qquad \rightarrow \qquad \lbrack \hat{\psi}_{1}({\bf r}),%
\hat{\pi}_{1}({\bf r^{\prime }})]\ =\ i\hbar \delta ({\bf r}-{\bf r^{\prime }%
})\ .  \label{qscheme2}
\end{equation}%
But modulo the constraints this is consistent with Eq. (\ref{comm3}). Now,
in contrast with the treatment of Schiff, we can impose $\hat{\psi}_{2}\ =\ 
\hat{\pi}_{2}\ =\ 0$, because in the framework of the constrained systems
these canonical variables are second class constraints and they should not
be turned into canonically conjugate operators \cite{Dirac2}.

\section{Concluding remarks}

We have reviewed how the Schr\"{o}dinger equation can be found from various
Hamiltonians, representing different stages of reduction. Hopefully our
treatment shed light on the many Hamiltonian formulations of the Schr\"{o}%
dinger system and their multiple interconnections. We have seen how the
construction of these Hamiltonians requires some artwork, like the addition
of appropriately chosen time derivative terms to the Lagrangian.

Alternatively the Schr\"{o}dinger field appears as a computationally simple
example for constrained systems. By employing the characteristic toolchest,
we have found the Hamiltonian and the Schr\"{o}dinger equation via the
Dirac-Bergmann algorithm and the consistency requirement, respectively. We
have achieved the reduction to the physical degrees of freedom by suitable
canonical transformations. The canonical equation in the reduced phase space
is again the Schr\"{o}dinger equation.

We have shown how second quantization of the complex Schr\"{o}dinger field by
turning the Dirac bracket (or the equivalent fundamental bracket in the
Faddeev-Jackiw approach) into commutators on the one side avoids
interpretational difficulties, on the other side leads to the same quantum
theory, which emerges from quantization on the reduced phase space.

The Sch\"{o}dinger equation completes the list of famous equations of modern
physics, like Maxwell and Einstein equations, not covered by the ''usual''
variational treatments. The exceptions turn out to be rather generic. However,
there is a major difference: the Schr\"{o}dinger constraints are second
class as opposed to the first class constraints of electrodynamics and
general relativity. Following the equivalent Faddeev-Jackiw approach, there
are no constraints at all. This latter approach with uncontested simplicity
yields the correct fundamental bracket for the Schr\"{o}dinger system, but
the role of the previously found Hamiltonians is revealed only by the Dirac
method.

\section{Acknowledgments}

The author is grateful to Karel Kucha\v r for his constructive criticism on
an early version of this paper, to Mih\'aly Benedict for bringing into his
attention Cohen-Tannoudji's approach and to J\'anos Polonyi for his
encouragement to pursue the topic. This work has been completed under the
support of the Zolt\'an Magyary Fellowship.

\end{document}